\documentclass[a4paper,11pt]{article}
\pdfoutput=1 

\usepackage{jcappub} 
\usepackage[T1]{fontenc} 

\title{\boldmath Decoherence effect in neutrinos produced in micro-quasar jets} 

\author[a,b]{M. E. Mosquera,}
\author[b,1]{O. Civitarese\note{Corresponding author.}}

\affiliation[a]{Facultad de Ciencias Astron\'{o}micas y Geof\'{\i}sicas, Universidad Nacional de La Plata, \\ Paseo del Bosque, (1900) La Plata, Argentina}
\affiliation[b]{Department of Physics, University of La Plata, \\ c.c. 67 (1900), La Plata, Argentina}

\emailAdd{mmosquera@fcaglp.unlp.edu.ar}
\emailAdd{osvaldo.civitarese@fisica.unlp.edu.ar}


\abstract{
We study the effect of decoherence upon the neutrino spectra produced in micro-quasar jets. In order to analyse the precession of the polarization vector of neutrinos we have calculated its time evolution by solving the corresponding equations of motion, and by assuming two different scenarios, namely: (i) the mixing between two active neutrinos, and (ii) the mixing between  one active and one sterile neutrino. We have found that for the case with two active neutrinos and large values of the neutrino-neutrino interactions the onset of decoherence is not manifest. For the active-sterile scheme decoherence becomes manifest
 if the strength of the neutrino-neutrino interactions ($\mu$) and the ratio between the square-mass difference and the energy ($\omega=\frac{\delta m^2}{2 E}$) satisfy the relation   $\frac{\mu}{w_{E=E_{\rm min}}}>0.1$.
}
\begin{document}
\maketitle
\flushbottom

\section{Introduction}
\label{Intro}
The study of neutrino's related processes in astroparticle physics is a subject of utmost importance since it is strongly connected with crucial aspects of particle physics \cite{mohapatra06}.  The achievements in the field, both theoretically  and experimentally  are impressive:
the values of the neutrino-flavor oscillations parameters have been determined \cite{forero14,gonzalez14,capozzi14,capozzi16} and various scenarios for the mass hierarchy have been proposed and constrained experimentally \cite{qian15}. Lately, the importance of neutrino-neutrino interactions in star evolution has been emphasized \cite{tamborra14a,tamborra14b}, particularly in dealing with supernovae's dynamics. The neutrinos are indeed very peculiar particles, since they can travel enormous distances without being severely affected by local interactions. However, their quantum nature should manifests in phenomena like decoherence \cite{schlosshauer04,zurek81,zurek82}. As pointed out in Ref. \cite{raffelt10}, the onset of decoherence may affect strongly the density and energy-momentum distribution of neutrinos produced in distant sources.

As it is well known from elementary quantum mechanics, pure states may evolve into mixed states due to interactions with the background \cite{schlosshauer04,zurek81}. If this is the case with neutrinos produced in supernovae explosions or in other astrophysical events the information about oscillation parameters, masses, etc, may be depending about the presence of decoherence. In this paper we focus on neutrinos produced from various reactions which take place in a micro-quasar. By modelling their spectra and initial densities we are able to follow their evolution in time and determine conditions for the appearance of decoherence. We have followed the formalism of \cite{christiansen06,romero03} and adapted it to calculate the pattern of decoherence in the time evolution of micro-quasar's neutrinos. We have found a dependence of the decoherence pattern with the mixing scheme of the neutrinos and with the neutrino-neutrino interactions. We have considered two cases, that is a) the mixing between active neutrinos and b) the mixing between active and sterile neutrinos.

The work is organized as follows. In Section \ref{deco} we present the formalism which we have developed to calculate the time evolution of the neutrino spectra and in Section \ref{espectra} we used it to obtain the neutrino spectrum in a micro-quasar jet. The results of the calculations are presented in Section \ref{resultados}. Our conclusions are drawn in Section \ref{conclusion}.

\section{Formalism}
\label{deco}

The time evolution of the occupation number of neutrinos is governed by the equation of motion \cite{raffelt10}
\begin{eqnarray}
\imath \dot{\rho_f}&=&\left[\frac{M^2 c^4}{2E \hbar}+ \sqrt{2}G_F \,\rho, \, \rho_f \,\,  \right] \, ,
\end{eqnarray}
where the squared brackets reads for the commutator and $\frac{M^2 c^4}{2E\hbar}$ is the mass-squared matrix in the flavour basis. The quantity $\rho_f$ is the density matrix in the flavour basis.

Following Ref. \cite{raffelt10}, one can write the mass matrix and the matrix $\rho_f$ in terms of Pauli matrices. For two-neutrino mass eigenstates they are written
\begin{eqnarray}
\frac{M^2}{2E\hbar}&=&\frac{1}{2} {\rm tr}\left(\frac{M^2}{2E\hbar}\right)I + \frac{1}{2} w \bar{B} \cdot \bar{\sigma} \, \, \, , \\
\rho_f &=&\frac{1}{2} {\rm tr}\left(\rho_f\right) I+ \frac{1}{2} \bar{P}_f\cdot \bar{\sigma} \, \, \, .
\end{eqnarray}
In the previous equation $w=\frac{\delta m^2}{2E\hbar}$ and $\delta m^2$ is the mass-squared difference between the mass eigenstates and $\bar{B}$ is an unitary vector which fixes the orientation of the background. $\bar{P}_f$ is the polarization vector in the flavor basis.
The equation of motion is then re-written as
\begin{eqnarray}
\label{pw}
\frac{\partial {\bar{P}_w}}{\partial t}&=& \left( w \bar{B}+\mu \bar{P}\right) \times \bar{P}_w \, \, \, ,
\end{eqnarray}
where $\mu$ stands for the neutrino-neutrino interaction, and 
\begin{eqnarray}
\label{p}
\bar{P}&=& \int \bar{P}_w \, dw \, \, \, ,
\end{eqnarray}
is the total (or global) polarizarion vector.
 
 One can perform a rotation in order to set this direction as the z-axis in flavour space, then the initial condition for $\bar{P}_w$, is given by
\begin{eqnarray}
\label{pw0}
\bar{P}_w(0)&=&\left(
\begin{array}{c}
\sin 2\theta \\
0\\
\cos 2 \theta
\end{array}
\right)g(w) \,\, \, ,
\end{eqnarray}
where $g(w)=A(g_\mu-g_e)$, $g_e$ and $g_\mu$ are the electron and muon-neutrino spectral functions, $A$ is a normalization constant and $\theta$ is the neutrino mixing angle. 

The order parameter that measures coherence is defined as the ratio between the modulus of the perpendicular polarization vector at time $t$ and it at time $t=0$, that is:
\begin{eqnarray}
\label{rtheta}
R_\theta(t)&=&\frac{\left|\bar{P}_\perp (t)\right|}{\left|\bar{P}_\perp (0)\right|}
\end{eqnarray}
where $\bar{P}_\perp (t) = \bar{P}-\left(\bar{P} \cdot \bar{B}\right) \bar{B}$. The average in angles can be computed as
\begin{eqnarray}
R(t)&=&\frac{\int R_\theta(t) \, {\rm d}\theta}{\int \, {\rm d}\theta}
\end{eqnarray}

\section{Neutrino spectra}
\label{espectra}

\subsection{Gaussian spectrum}

Following Ref. \cite{raffelt10} and as toy models we shall use two different Gaussian-like spectra
\begin{eqnarray}
g_{w1}(w)&=& \frac{1}{2\sqrt{2\pi}}e^{-(w-5)^2/2} + \frac{1}{2\sqrt{2\pi}}e^{-(w+5)^2/2}\, \, \, , \nonumber \\
g_{w2}(w)&=& \frac{1}{2\sqrt{2\pi}}e^{-(w-1)^2/2} + \frac{1}{2\sqrt{2\pi}}e^{-(w+1)^2/2}\, \, \, ,
\end{eqnarray}
that is two-non-overlapping Gaussian distributions, $g_{w1}$, and two-overlapping Gaussian functions, $g_{w2}$, respectively.

\subsection{Jet's neutrino spectrum in micro-quasar from $p\gamma$ and $pp$ interactions}
\label{jetall}

In order to compute the neutrino spectrum produced in a micro-quasar, we follow Ref. \cite{reynoso09}. We have assumed a compact object with an accretion disk and a perpendicular jet with a half-opening angle $\xi$. The injection point is located at $z_0$. In Table \ref{all} we show the parameters used in the calculation. The differential equation that gives the density of the particles, $N(E)$, in a micro-quasar jet is
\begin{eqnarray}\label{diffeq}
\frac{\partial \left(N(E) b(E)\right)}{\partial E} + \left(t^{-1}_{dec}+t^{-1}_{esc}\right) N(E)&=& Q(E)\, \, \, ,
\end{eqnarray}
where $b(E)= -E t^{-1}_{loss}$, $Q(E)$ is the particle injection, $t^{-1}_{dec}$ and $t^{-1}_{esc}$ are the decay and escape rate respectively. The rate $E t^{-1}_{loss}$ is the sum of the cooling rates, that is $t^{-1}_{syn}+t^{-1}_{ad}+t^{-1}_{pp}+t^{-1}_{p\gamma}$ for protons, $t^{-1}_{syn}+t^{-1}_{ad}+t^{-1}_{\pi p}+t^{-1}_{\pi\gamma}$ for pions and $t^{-1}_{syn}+t^{-1}_{ad}+t^{-1}_{ic}$ for muons, respectively. The other quantities needed to evalute the cooling rates are: $t^{-1}_{syn}$ which is rate of emission of synchrotron radiation, $t^{-1}_{ad}$ which stands for the adiabatic cooling, $t^{-1}_{pp}$ which is the $pp$ collision rate, $t^{-1}_{p\gamma}$ which gives the rate of the interaction between protons and synchrotron photons. Finally, $t^{-1}_{\pi p}$ and $t^{-1}_{\pi\gamma}$ represent the proton-pion and pion-photon interaction rates and $t^{-1}_{ic}$ is the loss-rate of the inverse Compton interactions \cite{reynoso09}. The solution of  equation (\ref{diffeq}) is 
\begin{eqnarray}\label{nze}
N(z,E)&=& \frac{1}{\left|b(z,E)\right|} \int_{E}^{E_p^{max}}  Q(z,x) e^{-\tau(E,x,z)}{\rm d}x\, 
\end{eqnarray}
where $E>1.2 \, {\rm GeV}$.The proton maximum energy $E_p^{max} = 5 \times 10^6 \, {\rm GeV}$ is obtained through the assumption that the acceleration rate is equal to the loss rate at the initial high of the jet and
\begin{eqnarray}
\tau(E,x,z)&=&\int_E^x {\rm d}y  \frac{t^{-1}_{dec}(y)+t^{-1}_{esc}(z)}{\left|b(z,y)\right|}\, \, \, . 
\end{eqnarray}
\begin{table}[!h]
\renewcommand{\arraystretch}{1.3}
\begin{center}
\caption{Micro-quasar parameters \cite{reynoso09}.} \label{all}
\begin{tabular}{|c|c|c|}
\hline Parameter & Value\\ \hline
Jet power &$5\times 10^{33} \, {\rm J \, s}^{-1}$ \\ \hline
Initial jet's high & $10^{5} \, {\rm m}$\\ \hline 
Lorentz factor &$1.25$\\ \hline 
Relativistic particles &$0.1$\\ \hline
Hadron-to-lepton ratio &$100$\\ \hline
Half opening angle & $0.087$\\ \hline
\end{tabular}
\end{center}
\end{table}

\subsubsection{Proton injection}

The proton injection, $Q(z,E)$ of Eq.(\ref{nze}) is \cite{reynoso09}
\begin{eqnarray}
 Q(z,E)&=&Q_0^p \left(\frac{z_0}{z}\right)^3 \Gamma^{-1}\left(E-\beta \sqrt{E^2-m_p^2 c^4}\cos\theta\right)^{-2} 
\left(\Gamma -\cos \theta \frac{E \beta}{\sqrt{E^2-m_p^2 c^4}}\right)\, \, \, . 
\end{eqnarray}
$Q_0^p$ is a constant to be determined from the luminosity, $\Gamma$ is the Lorentz factor, $\theta$ is the observation angle and $\beta$ is related to $\Gamma$ (see \cite{reynoso09} for details). In this case $t^{-1}_{dec}=0$.

\subsubsection{Pion injection}

The pion injection is $Q_\pi(E)=Q_\pi^{pp}(E)+Q_\pi^{p\gamma}(E)$, where $Q_\pi^{pp}(E)$ and $Q_\pi^{p\gamma}(E)$ are the injection terms resulting from the pion production due to proton-proton and proton-photon interactions, respectively.
The proton-proton injection is calculated as
\begin{eqnarray}
Q_\pi^{pp}(z,E)= n_p(z) c\int_E^{E_p^{max}} \hskip -0.8cm N_p(z,E_p) F_\pi \left(\frac{E}{E_p}, E_p\right) 
\sigma_{pp} \left(E_p\right) \frac{{\rm d} E_p}{E_p} . \nonumber
\end{eqnarray}
In the previous equation, $n_p$ is the density of cold particles, $N_p(z,E_p)$ is the proton density, $ \sigma_{pp}$ is the cross section and $F_\pi$ is the pion's distribution produced per $pp$ collisions (see Ref. \cite{reynoso09,kelner06})

The proton-photon production is given by the expression
\begin{eqnarray}
Q_\pi^{p\gamma}(z,E)&=& 5  N_p(z,5 E,\theta) \omega_{p\gamma} \left(z,5 E\right) \mathcal{N}_{\pi}\left(z,5 E\right) \, \, ,
\end{eqnarray}
where $\omega_{p\gamma}$ is the collision-frequency and $\mathcal{N}_{\pi}$ the mean number of positive and negative pions.

\subsubsection{Muon injection}

The pion decay produces muons, therefore the muon injection is
\begin{eqnarray}
Q_\mu^{L^-; \, R^+}(z,E) &=& \int_E^{E_p^{max}} t^{-1}_{dec\, \pi}\left(E_\pi\right) N_\pi\left(z,E_\pi\right)  
\frac{{\rm d}n_{\pi^- \rightarrow \mu_L^-}}{{\rm d}E}\left(E,E_\pi\right) {\rm d}E_\pi \, \, \, , \nonumber\\ 
Q_\mu^{R^-; \, L^+}(z,E)&=&\int_E^{E_p^{max}} t^{-1}_{dec\, \pi}\left(E_\pi\right) N_\pi\left(z,E_\pi\right) 
\frac{{\rm d}n_{\pi^- \rightarrow \mu_R^+}}{{\rm d}E}\left(E,E_\pi\right) {\rm d}E_\pi   \, \, \, .  
\end{eqnarray}
In the previous expression $t^{-1}_{dec\, \pi}$ is the pion decay rate, $N_\pi\left(z,E_\pi\right)$ the density of pions and $\frac{{\rm d}n_{\pi^- \rightarrow \mu_L^-}}{{\rm d}E}$ and $\frac{{\rm d}n_{\pi^- \rightarrow \mu_R^+}}{{\rm d}E}$ are the decay rates of left-handed and right-handed muons, respectively \cite{lipari07}.

\subsubsection{Neutrino injection}

The neutrino production due to pion decay can be written as
\begin{eqnarray}
Q_{\pi\rightarrow \nu}(z,E) &=& \int_E^{E_p^{max}}  t^{-1}_{dec\, \pi}\left(E_\pi\right) N_\pi\left(z,E_\pi\right) 
\Theta\left(1-\frac{E}{E_\pi} -\left(\frac{m_\mu}{m_\pi}\right)^2 \right)  
\left(1-\left(\frac{m_\mu}{m_\pi}\right)^2 \right)^{-1} \frac{{\rm d}E_\pi}{E_\pi} ,\nonumber\\
\end{eqnarray}
The neutrino injection due to the muon decay is
\begin{eqnarray}
Q_{\mu\rightarrow \nu}(z,E) &=& \sum_{i=1}^4\int_E^{E_p^{max}} \hskip -0.8cm t^{-1}_{dec\, \mu}\left(E_\mu\right)  N_{\mu_i}\left(z,E_\mu,\right) 
y\left(\frac{E}{E_\mu} \right) \frac{{\rm d}E_\mu}{E_\mu}\, ,
\end{eqnarray}
where $t^{-1}_{dec\, \mu}$ is the muon decay rate, $N_{\mu_i}\left(z,E_\mu,\right)$ the muon density and $y(x)$ is a polynomial function  \cite{reynoso09}.

The neutrino spectral function can be calculated  by performing the integral in the jet volume
\begin{eqnarray}
g_\mu\left(E_\nu\right)&=&\int {\rm d} V  \frac{Q_{\mu\rightarrow \nu}(z,E)+Q_{\pi\rightarrow \nu}(z,E)}{t_{esc}^{-1}(z)}\nonumber \\
g_e\left(E_\nu\right)&=&\int {\rm d} V  \frac{Q_{\mu\rightarrow \nu}(z,E)}{t_{esc}^{-1}(z)}
\end{eqnarray}

\subsection{Jet's neutrino spectrum in windy micro-quasar}
\label{windy}

In order to calculate the neutrino density as a function of the energy, we have assumed that a binary system formed by one high-mass primary star sub-rounded by a disk and a compact object describing a Kepler-orbit (see Table \ref{vientos}) and followed the analysis presented in Ref.\cite{christiansen06}. The jet of relativistic particles produced by the compact object is considered to be cone perpendicular to the accretion-disk plane (or orbital plane). We have used the wind velocity model \cite{romero03}
\begin{eqnarray}
v(r_w)=v_\infty \left(\frac{R_\star}{r_w}\right)^{1.2} \, \, \, ,
\end{eqnarray}
where $r_w$ is the radial coordinate from the center of the star, $R_\star$ is the star radius, $v_\infty$ is the terminal velocity of the wind. The mass density of the wind is obtained from the continuity equation \cite{raffelt10}.
\begin{table}[!h]
\renewcommand{\arraystretch}{1.3}
\begin{center}
\caption{Windy micro-quasar parameters (from Ref.\cite{christiansen06}).} \label{vientos}
\begin{tabular}{|c|c|c|}
\hline Parameter & Value\\ \hline
$M_\star$ & $10 \, R_\odot$ \\ \hline
$M_{\rm compact \, object}$ & $1.4 \, R_\odot$ \\ \hline
$R_\star$ & $ 10\, R_\odot$\\ \hline
Period & $26.496 \, {\rm days}$  \\ \hline
Eccentricity & $0.72$  \\ \hline
Initial orbital phase &$ 0.261799$  \\ \hline
$\rho_0$ & $10^{-8} \, {\rm kg \, m}^{-3}$\\ \hline
$v_\infty$& $5 \, {\rm km \, s}^{-1}$\\ \hline 
Initial jet's high & $10^{5} \, {\rm m}$\\ \hline 
Initial jet's radius & $10^{4} \, {\rm m}$\\ \hline 
Proton spectrum power law&$2.2$\\ \hline 
Lorentz factor &$1.25$\\ \hline 
\end{tabular}
\end{center}
\end{table}

Following Ref. \cite{romero03} one can write the proton spectrum in the jet frame and the accretion rate due to the wind to obtain the proton flux in the observer frame \cite{romero03,christiansen06}. These protons interact with target protons of the wind {\it{via}} the reaction
\begin{eqnarray}
p+p&\rightarrow& p+p +\xi_{\pi^0}(E_P) \pi^0+\xi_{\pi}(E_P) (\pi^+ + \pi^-) \, \, \, , 
\end{eqnarray}
where $\xi_{\pi^0}(E_P)$ and $\xi_{\pi}(E_P)$ are the multiplicities for neutral and charged pions, given by  \cite{christiansen06}
\begin{eqnarray}
\xi_{\pi^0}(E_p)&=& 1.1 \left(\frac{E_p}{\rm GeV}\right)^{1/4} \, \, \, ,  \nonumber\\
\xi_\pi(E_p) &=& \left(\frac{E_p}{\rm GeV} -1.22\right)^{1/5}\, \, \, . 
\end{eqnarray}
The proton ($E_p$) and photon ($E_\gamma$) energies are related by $E_p= 6 k^{-1} \xi_{\pi^0}(E_p) E_\gamma$, where $k=0.5$ is the inelasticity coefficient.

From the energy conservation one can obtain the neutrino intensity produced by pion- and muon-decay \cite{stecker96,alvarez02}
\begin{eqnarray}
\int_{E_\gamma^{min}}^{E_\gamma^{max}}{\rm d} E_\gamma \frac{{\rm d}N_\gamma}{{\rm d}E_\gamma} E_\gamma &=&\Delta \int_{E_\nu^{min}}^{E_\nu^{max}}{\rm d} E_\nu \frac{{\rm d}N_\nu}{{\rm d}E_\nu} E_\nu \, \, \, ,
\end{eqnarray}
where $E_\gamma^{min}$ and $E_\gamma^{max}$ are the minimum and maximum energies of photons resulting from hadrons, and $E_\nu^{min}$ and $E_\nu^{max}$ are the corresponding minimum and maximum energy of the neutrinos, and $\Delta=1$ \cite{alvarez02}. The neutrino energy is related to the photon energy by $E_{\nu}= \frac{1}{2} E_\gamma$, leading to \cite{ginzburg64}
\begin{eqnarray}
E_{\nu}&=& \frac{k}{12 \xi_\pi(E_p)}E_p \, \, \, .
\end{eqnarray}

The maximum neutrino energy is determined by the maximum energy acquired by the accelerated protons, which is related to the magnetic field $B$. The magnetic field is calculated by assuming equipartition between the magnetic field energy and the kinetic energy of the jet \cite{christiansen06}. The maximum energy of the protons is
\begin{eqnarray}
E_p(\psi)&=&e R(z_0) B(\psi,z_0) \, \, \, .
\end{eqnarray}
Note that $0.5\, {\rm GeV}<E_p(\psi)<2.8 \times 10^4 \, {\rm GeV}$.

The muon-neutrino density computed as 
\begin{eqnarray}
G_\mu(E_\nu, \psi)&=& \frac{4f_p}{m_p}\int {\rm d} V \frac{ \rho_w(\psi;z,\delta,\phi) q_\gamma(\psi;2E_\nu,z,\theta)}{t_{esc}^{-1}(z)} \,,
\end{eqnarray}
where $f_p=0.1$ takes into account particle-rejection from the boundary \cite{romero05b}, $\rho_w(\psi;z,\delta,\phi)$ is the wind mass density,  $q_\gamma(\psi;2E_\nu,z,\theta)$ stands for the gamma-ray emissivity \cite{aharonian96} and $t_{esc}^{-1}(z)=\frac{c}{z_m-z}$ is the inverse of the neutrino's escape time. The integral is performed in the jet's volume.

The muon neutrino spectrum is then calculated as
\begin{eqnarray}
g_\mu(w)&=&\frac{1}{2\pi}\int_0^{2\pi} {\rm d}\psi \, G_\mu(E_\nu, \psi)\,\,\, .
\end{eqnarray}
The electron-neutrino spectrum can be determined by repeating the same arguments. 

\section{Results}
\label{resultados}

We have considered two different neutrino's scenario, that is: (i) two-active neutrinos and one active and (ii) one sterile neutrino, to compute the order parameter. 
The neutrino-mixing parameters for the first scenario are $\delta m^2=7.53 \times 10^{-5} \, {\rm eV}^2$ and $\sin^2 \theta=0.307$ \cite{mixing}. In this case, we have computed both the electron- and muon-neutrino spectral functions produced in the micro-quasar and used them as initial condition (see Eq. \ref{pw0}) to compute the time dependence of the polarization vector and the order parameter. For the active-sterile neutrino mixing parameters we have used $\delta m^2=1 \, {\rm eV}^2$ and $\sin^2 \theta=0.1$. 
As initial condition for the sterile neutrino sector we assume that 
sterile neutrino are not produced in the micro-quasar jet.

\subsection{Gaussian spectra}

\begin{figure*}[h!]
\begin{center}
\includegraphics[scale=0.28,angle=-90]{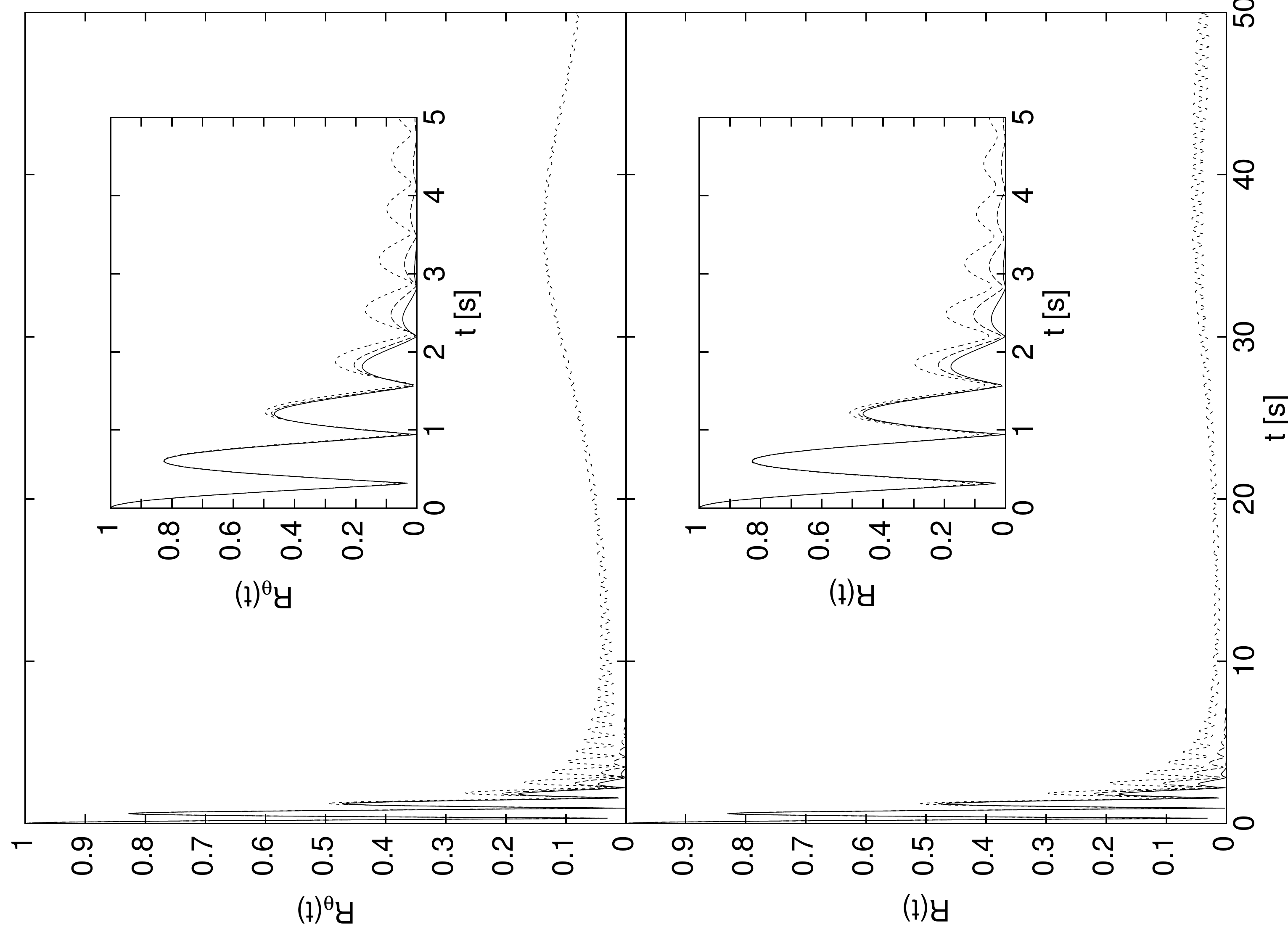}
\includegraphics[scale=0.28,angle=-90]{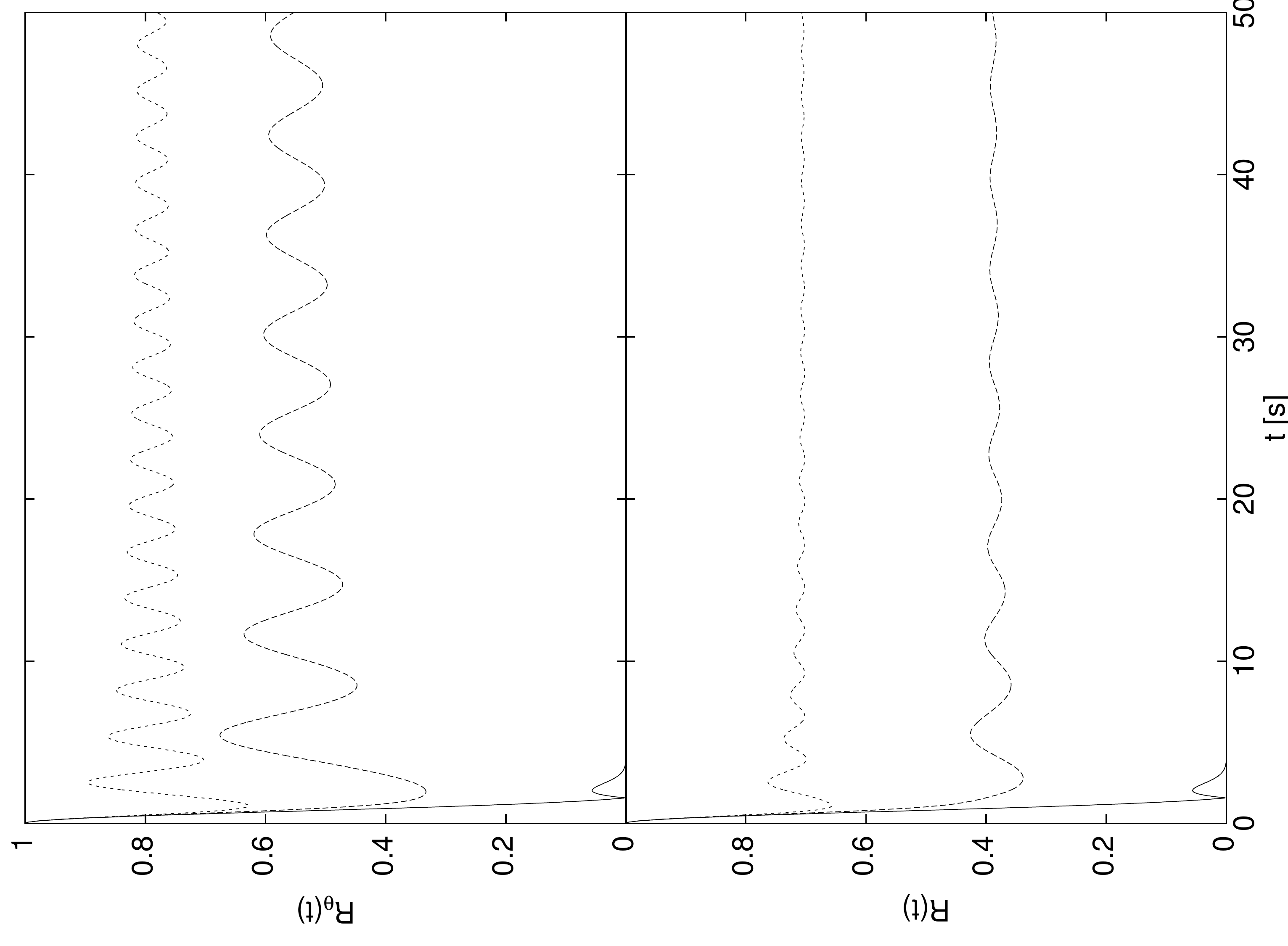}
\end{center}
\caption{{\it Left figure:} order parameter as a function of the time, for two-separated Gaussian spectra. Top figure: order parameter using the active-active mixing angle; bottom figure: mean value of the order parameter. Solid line: $\mu = 0 \, {\rm s}^{-1}$; dashed line: $\mu = 1 \, {\rm s}^{-1}$; dotted line: $\mu = 1.8 \, {\rm s}^{-1}$. {\it Right figure:} order parameter as a function of the time, for two-overlapping Gaussian spectra. Top figure: order parameter using the active-active mixing angle; bottom figure: mean value of the order parameter. Solid line: $\mu = 0 \, {\rm s}^{-1}$; dashed line: $\mu = 2 \, {\rm s}^{-1}$; dotted line: $\mu = 3 \, {\rm s}^{-1}$.} \label{r-2-gauss}
\end{figure*}

As toy model we have computed the order parameter for two-active neutrino's Gaussian spectrum. In Figure \ref{r-2-gauss} we show the order parameter for the two-Gaussian spectra at the mixing angle (top figure) and its mean value (bottom figure), as a function of the time and for different values of $\mu$. In absence of neutrino-neutrino interactions, that is $\mu=0$, the order parameter shrinks to zero as well as its mean value. When the neutrino-neutrino interaction is activated $|\bar{P}|$ decreases its value but oscillates around a non-zero value. The larger the value of the interaction the larger is $|\bar{P}|$. The mean value of the order parameter reaches a smaller average value earlier than the one calculated with a fixed mixing angle.

The results for the case of two-overlapping Gaussian as initial condition of the neutrino 
spectrum, shown at the right inset of Figure \ref{r-2-gauss}, are similar to the ones obtained in Ref. \cite{raffelt10} for a Gaussian spectrum.

\subsection{Two-active neutrinos}

\begin{figure*}[!h]
\begin{center}
\includegraphics[scale=0.28,angle=-90]{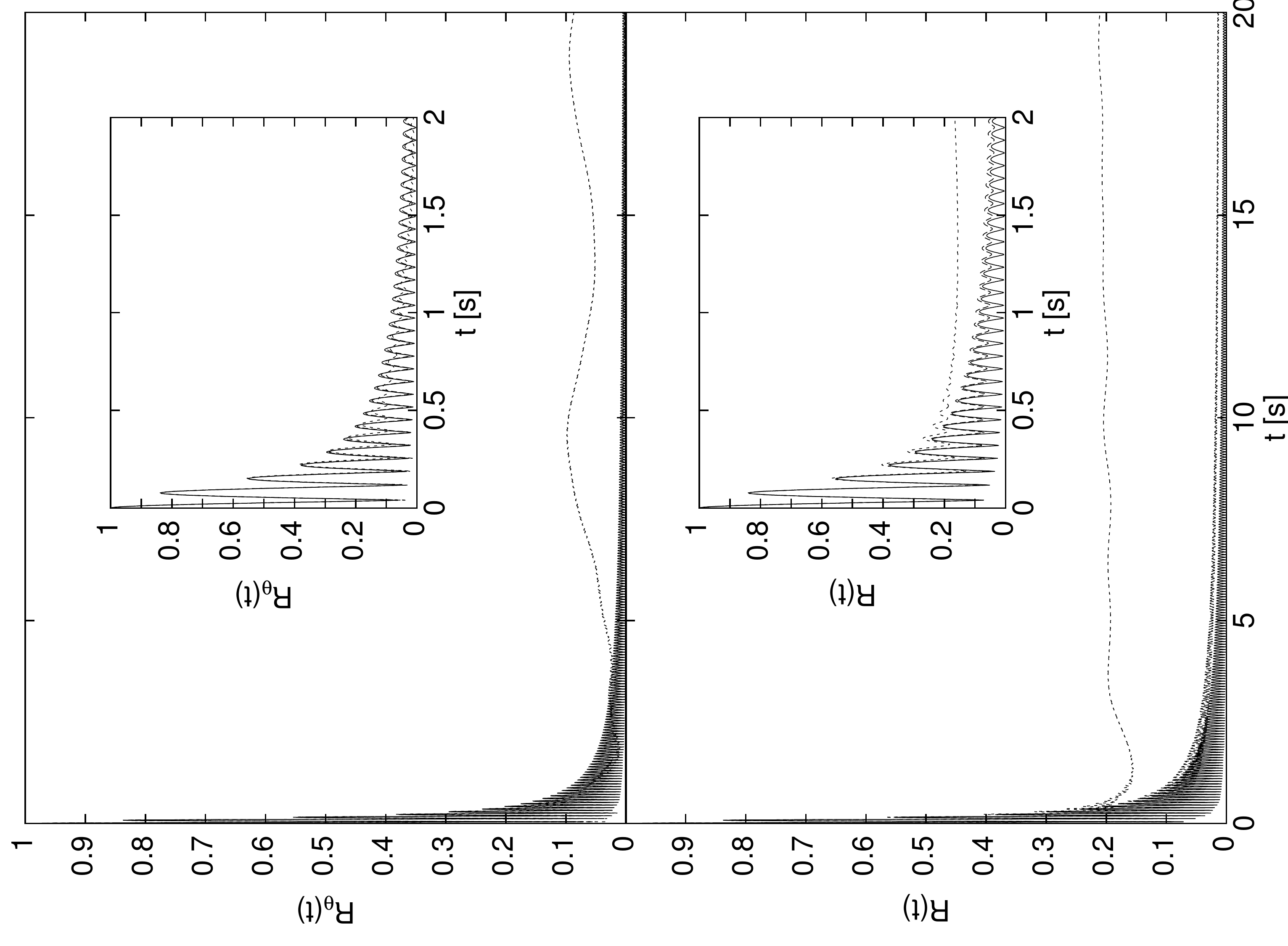}
\includegraphics[scale=0.28,angle=-90]{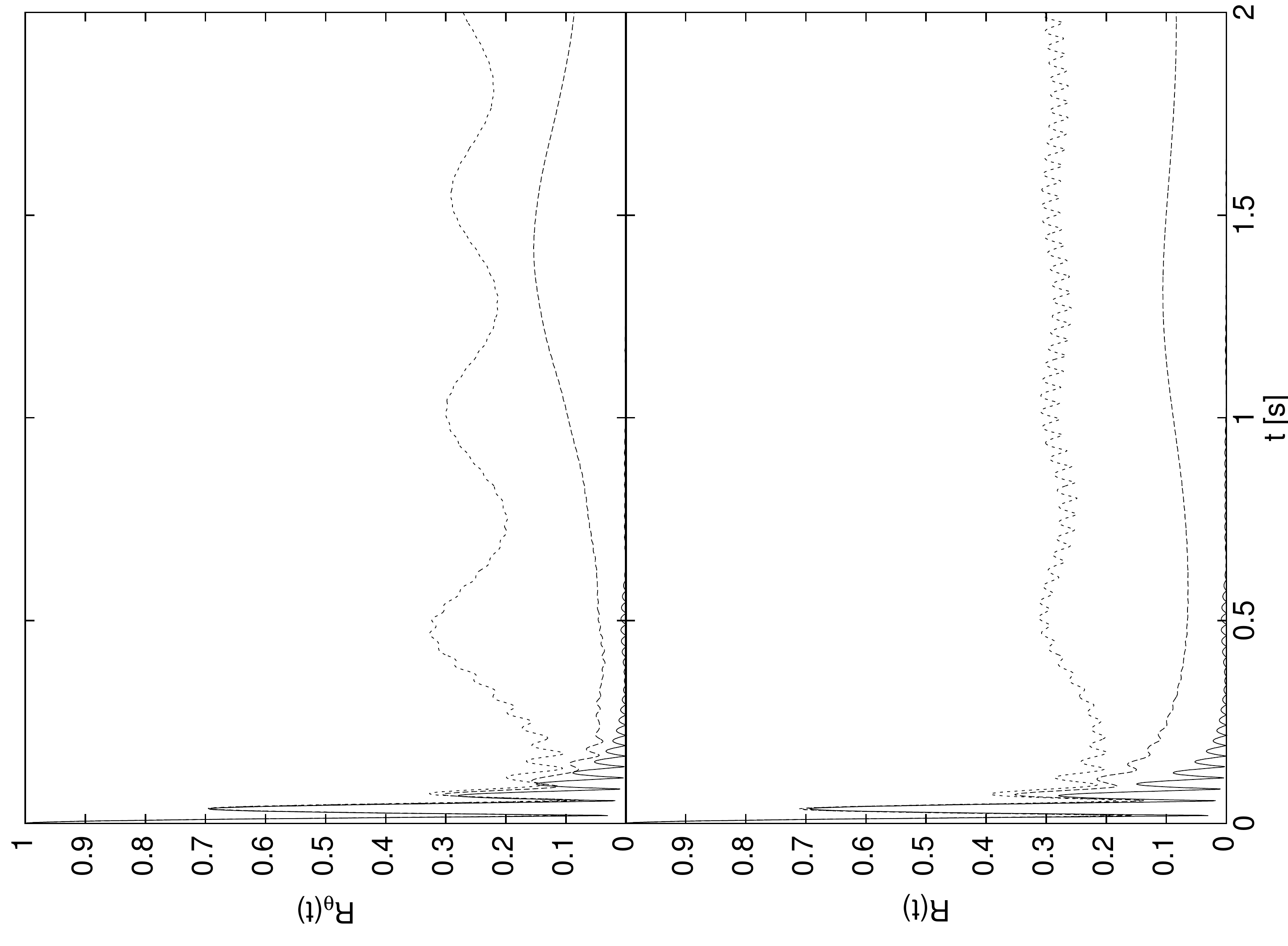}
\end{center}
\caption{{\it Left figure:} order parameter as a function of time calculated from the emission of neutrinos in a micro-quasar's jet and two-active neutrino. Top figure: order parameter using the active-active mixing angle; bottom figure: mean value of the order parameter. Solid line: $\mu = 0 \, {\rm s}^{-1}$; dashed line: $\mu/w_{max}=0.08$; dotted line $\mu/w_{max}=0.2$. 
{\it Right figure:} order parameter as a function of time calculated from the neutrino spectrum in a windy micro-quasar and two-active neutrino. Top figure: order parameter using the active-active mixing angle; bottom figure: mean value of the order parameter. Solid line: $\mu = 0 \, {\rm s}^{-1}$; dashed line: $\mu/w_{max}=0.32$; dotted line $\mu/w_{max}=0.40$.}
\label{r-active}
\end{figure*}

In this section we present the results for the order parameter calculated by using the neutrino spectral function
as described in Section \ref{jetall} and in Section \ref{windy}. The results for the first case are shown in the left inset of Figure \ref{r-active}. They are similar to the ones obtained using as initial condition the two-non-overlapping Gaussian functions. The length of the vector $\bar{P}$ is reduced to zero for the non-interacting case and for small values of $\mu$. However, for larger values of the neutrino-neutrino interaction, the polarization vector oscillates towards an asymptotic non-zero value.

The order parameter $R$ obtained using the neutrino spectra described in section \ref{windy} is shown in the right inset of Figure \ref{r-active}, for active-active neutrino scheme. The initial neutrino spectra can be modelled as two-non-overlapping Gaussian functions, that's why the results presented in this section are quite similar to the previous ones.

\subsection{Active-sterile neutrino}

\begin{figure*}[!ht]
\begin{center}
\includegraphics[scale=0.28,angle=-90]{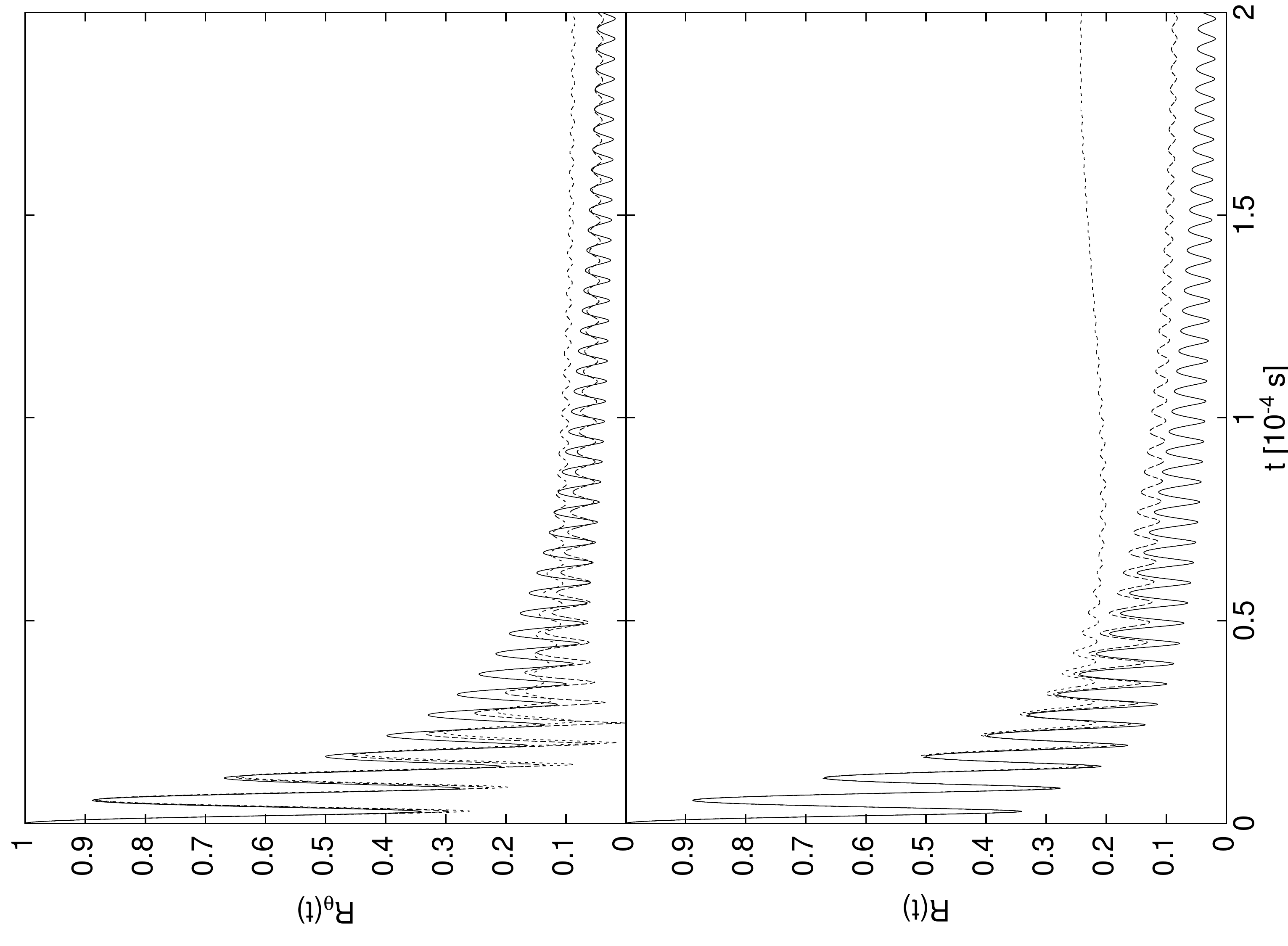}
\includegraphics[scale=0.28,angle=-90]{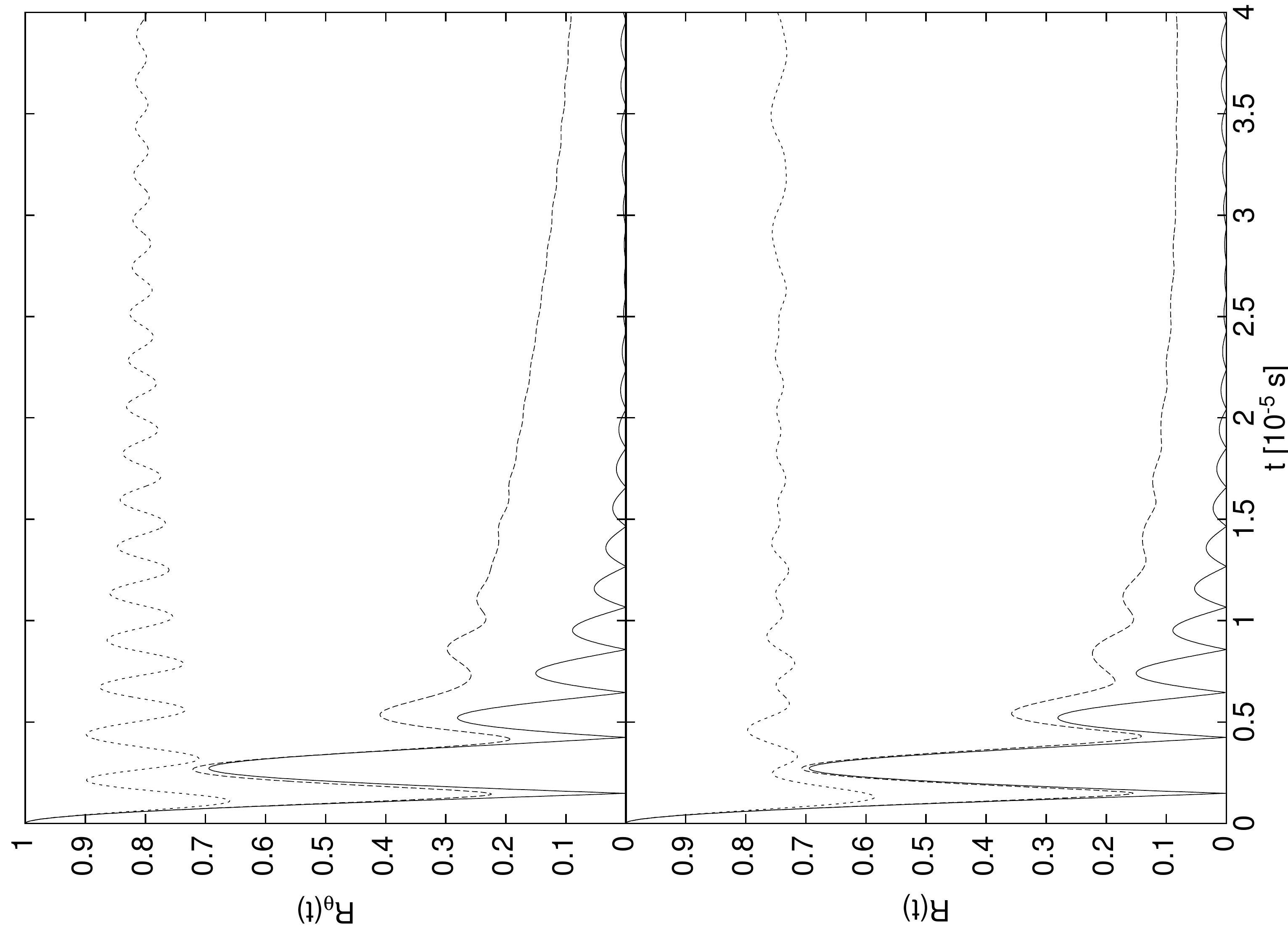}
\end{center}
\caption{{\it Left figure:}  order parameter as a function of the time calculated from the emission of neutrinos in a micro-quasar's jet and for active-sterile neutrino scheme. Top figure: order parameter using the active-active mixing angle; bottom figure: mean value of the order parameter. Solid line: $\mu = 0 \, {\rm s}^{-1}$; dashed line: $\mu/w_{max}=0.1$; dotted line $\mu/w_{max}=0.16$. 
{\it Right figure:} order parameter as a function of the time calculated from the neutrino spectrum in a windy micro-quasar and active-sterile  neutrino scheme. Top figure: order parameter using the active-active mixing angle; bottom figure: mean value of the order parameter. Solid line: $\mu = 0 \, {\rm s}^{-1}$; dashed line: $\mu/w_{max}=0.33$; dotted line $\mu/w_{max}=1.25$.}
\label{r-esteriles}
\end{figure*}

For the active-sterile neutrino, the results are shown in Figure \ref{r-esteriles}. The results displayed in the left inset of this figure have been obtained by applying the formalism described in section \ref{jetall}. As seen from the curves, the length of the vector $\bar{P}$ is reduced to zero for the non-interacting case but it does not vanishes for larger values of the neutrino-neutrino interaction. 

For the windy micro-quasar jet formalism one can see that, for large values of the parameter $\mu$, the length of the polarization vector is reduced by a $20\%$, as shown in the right inset of Figure \ref{r-esteriles}. Meanwhile, for small values of the interaction, the length of the vector $\bar{P}$ is depleted.

\section{Conclusions}
\label{conclusion}

In this work we have studied the effect of collective oscillations upon the neutrino spectral function, for neutrinos produced in micro quasar's jets, by applying the formalism developed in Refs. \cite{christiansen06,reynoso09}. Using active neutrinos as initial condition for the evolution of the polarization vector $\bar{P}$ we have calculated the order parameter $R_\theta$ as a function of the time.

For the case of neutrino's Gaussian spectra we have found that the polarization vector reduces its length to zero for small or null neutrino interactions, exhibiting a complete decoherence-pattern. If the neutrino density is large enough the length of the polarization vector becomes smaller than one and it oscillates around a non-zero asymptotic value. For non-overlapping Gaussians the reduction of the order parameter is faster and the neutrino density must increase in order to reduce the effects of decoherence.

The realistic electron- and muon-neutrino spectra produced in a micro quasar's jet are quite similar, for both formalisms
\cite{reynoso09, christiansen06}, to the two non-overlapping Gaussian spectrum. For this reason, the results are also quite similar to the ones obtained with the toy (Gaussian) models. The polarization vector is reduced to zero for small neutrino density while
 for larger values of the neutrino-neutrino interactions, the decoherence is not completed. This effect is quite noticeable, e.g. for $\mu=50 \, {\rm s}^{-1}$ ($\mu/w_{max}=0.40$), which results in a mean value of the order parameter equals to $0.3$ at large times.

When a sterile neutrino is oscillating with a light active-neutrino (electron-neutrino) the effects of the collective oscillations are noticeable at very small times, since the maximum value for the frequency $w$ is quite large due to the mass difference. In this case, for the electron-neutrino spectral function calculated in Section \ref{jetall}, the decoherence is almost complete for the mixing angle used, but the mean value of the order parameter is different from zero for $\mu/w_{max}>0.1$. For the case of a windy micro-quasar the polarization vector is reduced and this reduction is smaller if the neutrino density is larger.

Finally, the effect of decoherence is present in the neutrino flux produced in micro-quasar jets, with or without sterile neutrino. This effect is noticeable for small values of the neutrino density. For active-sterile neutrino mixing the effect becomes noticeable at  earlier times than for the mixing between active neutrinos.

\section*{{\bf Acknowledgments}}
This work was supported by a grant of the National Research Council of Argentina (CONICET), and by a research-grant of the National Agency for the Promotion  of Science and Technology (ANPCYT) of Argentina. The authors are members of the Scientific Research Career of the CONICET.

\bibliography{bibliografia}
\bibliographystyle{jhep}
\end{document}